\documentclass[fleqn,usenatbib]{mnras}

\usepackage{newtxtext,newtxmath}

\usepackage[T1]{fontenc}

\DeclareRobustCommand{\VAN}[3]{#2}
\let\VANthebibliography\thebibliography
\def\thebibliography{\DeclareRobustCommand{\VAN}[3]{##3}\VANthebibliography}

\usepackage{graphicx}	
\usepackage{amsmath}	
\usepackage{soul}
\usepackage[utf8]{inputenc}

\newcommand{\software}[1]{\mbox{\sc {#1}}}

\title[CD-27 2812]{Fundamental effective temperature measurements for eclipsing binary stars -- VIII. NIRPS spectroscopy of CD$-$27 2812}

\author[N. J. Adshead et al.]{
N. J. Adshead, 
P. F. L. Maxted,$^{\href{https://orcid.org/0000-0003-3794-1317}{\includegraphics[scale=0.5]{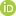}}}$\thanks{Corresponding author. E-mail: p.maxted@keele.ac.uk}
and A. Hahlin$^{\href{https://orcid.org/0000-0001-8365-8606}{\includegraphics[scale=0.5]{Images/orcid.jpg}}}$ \\
Astrophysics Group, Keele University, Staffordshire ST5 5BG, UK\\
\\
}
\date{Accepted XXX. Received YYY; in original form ZZZ}

\pubyear{2026}

\begin{document}
\label{firstpage}
\pagerange{\pageref{firstpage}--\pageref{lastpage}}
\maketitle

\begin{abstract}
There are very few M-dwarfs with accurate independent measurements of their mass, radius and effective temperature (T$_{\rm eff}$) that can be used to test stellar models for these low-mass stars. 
We aim to use high-resolution, near-infrared spectroscopy to measure the mass of M-dwarfs in eclipsing binary systems with solar-type stars and to measure the flux ratio between the two stars at near-infrared wavelengths. This information can then be combined with the analysis of the light curve, photometry, and the parallax to measure the mass, radius and T$_{\rm eff}$ for both stars. 
We have used the TESS light curve and spectra observed with the HARPS and NIRPS spectrographs to measure the following model-independent radii and masses for CD$-$27 2812, an F9\,V star in an eclipsing binary with a much fainter M~dwarf companion on a short near-circular orbit (P=7.8 d) :  $R_1 = 1.721  \pm 0.004  R_{\odot}$, $R_2 = 0.531  \pm 0.002 R_{\odot}$, $M_1 = 1.3597 \pm 0.0024  M_{\odot}$, and $M_2 = 0.5624 \pm 0.0006  M_{\odot}$
We show how the NIRPS spectra can be used to measure the flux ratio in the J and H bands.
This information, combined with published photometry and the Gaia DR3 parallax, leads to the following effective temperature measurements: $T_{\rm eff,1} = 6197 \pm 55$\,K, $T_{\rm eff,2} = 3770 \pm 28$\,K. 
This study demonstrates that it is now feasible to use eclipsing binaries to accurately measure T$_{\rm eff}$ for M-dwarf stars for which we also have independent mass and radius measurements.
\end{abstract}




\section{Introduction}

The discovery of planets orbiting Sun-like stars approximately 30 years ago \citep{1995Natur.378..355M, 2000ApJ...529L..45C} has led to the development of new instrumentation and data analysis techniques that can be used to discover and characterise these exoplanets \citep{2018exha.book.....P}. 
The improved access to efficient echelle spectrographs on many telescopes, the availability of high-quality light curves observed from space \citep{2021Univ....7..369S}, and the availability of software that can be used to analyse these data using sophisticated statistical methods has also been of huge benefit to stellar astrophysics. 
The developments go hand-in-hand because almost all measurements of an exoplanets properties are made relative to its host star, so accurate characterisation of exoplanets requires accurate estimates for their host star's fundamental parameters (mass, radius, composition, etc.).

The developments in instrumentation driven by exoplanet research now make it feasible to measure accurate masses and radii for stars in detached eclipsing binary systems (DEBS) with a precision of about 0.1\,per~cent in the best cases \citep{2020MNRAS.498..332M}.
These measurements can be combined with parallax measurements from the Gaia mission \citep{2016A&A...595A...1G,2023A&A...674A...1G} and photometry at ultraviolet, optical and near-infrared wavelengths from surveys to directly and accurately measure the effective temperature of stars in DEBS \citep{2020MNRAS.497.2899M}.
\citet{2022MNRAS.513.6042M} used near-infrared spectroscopy to demonstrate that direct mass, radius and effective temperature (T$_{\rm eff}$) measurements are feasible for solar-type stars eclipsed by very late M-dwarf companions despite the contrast ratio in the V-band being $\approx $ 8\,magnitudes.
Extensive monitoring of these eclipsing binaries with low mass companions (EBLM systems) to detect circumbinary planets has made it possible to make these direct mass, radius and T$_{\rm eff}$ measurements using echelle spectrographs operating at optical wavelengths \citep{2024MNRAS.530.2572S, 2024MNRAS.531.4577M, 2025MNRAS.541.2801B,2025MNRAS.544.4611M}.

EBLM systems with direct and accurate T$_{\rm eff}$ and $\log g$ measurements are ideal benchmark stars to test the accuracy of estimates for these quantities based on the analysis of the star's spectrum. 
They are within the magnitude range of recent large-scale spectroscopic surveys so they can be observed in the same way as other stars observed by these survey instruments as part of their routine operations. The small contribution of the M-dwarf to the total flux will have a negligible effect on the atmospheric parameters derived from the analysis of the optical spectrum. 
This enables the use of eclipsing binaries as calibrators for ``end-to-end'' test of the accuracy of parameters derived by the combination of these survey instruments plus their data processing and analysis pipelines.
However, direct mass measurement for EBLM systems are expensive in terms of the the amount and telescope time required so the sample of such stars is currently quite small. 
 
In order to increase the sample of benchmark stars, \citet{2023MNRAS.522.2683M} identified a sample of 20 DEBS with flux ratios $\approx 1$\,per~cent at optical wavelengths. 
These DEBS have narrow, total eclipses and little intrinsic photometric variability, and are moderately bright ($V \approx 10$), so accurate mass radius and  T$_{\rm eff}$ measurements for these stars is possible using much less telescope time than is required for EBLM systems.
This was demonstrated for the star HD~22064 by \citet{2023MNRAS.522.2683M} using 6 spectra observed at near-infrared wavelengths using a 2.5-m telescope  \citep{2020AJ....160..120J}.

There is estimated to be $\simeq$ 1 planet orbiting each low period-low mass star \citep{2025A&A...696A.101K}. With M-dwarfs constituting about 70\% of the stars found in the Galaxy \citep{2015ApJ...804...64M}, M-dwarfs are prime targets for identifying small Earth-like exoplanets in the stars habitable zone \citep{2019ApJ...871...63M}. 
Being able to determine the fundamental properties of these M-dwarfs is integral to determining the properties of their planets. 
Evolution models are an important part of determining the properties of stars, but it has been found that the models are less accurate for low mass stars than for Sun-like stars, e.g. the tend to underestimate the radius at a given mass \citep[``radius inflation problem'',][]{2013ApJ...776...87S}.
There is also significant variation in the predicted spectra of M-dwarfs from different stellar atmospheric models \citep{Iyer_2023,2026ApJ...997L...8G}.
Accurate fundamental data for M-dwarfs is needed to better understand these issues but has been difficult to obtain for this intrinsically faint stars.

In this study, we present our analysis of CD$-$27~2812, the first of the new benchmark stars identified by \cite{2023MNRAS.522.2683M} that we have analysed using new optical and near-infrared spectroscopy obtained with the HARPS \citep[High Accuracy Radial velocity Planet Searcher, ][]{2003Msngr.114...20M} and NIRPS \citep[Near Infra Red Planet Searcher,][]{2017Msngr.169...21B} spectrographs, respectively, on the ESO 3.6-m telescope.
CD$-$27~2812 is an F9\,V star with an M-dwarf companion star that transits the primary star once every 7.8 days. 
We demonstrate that we can accurately measure the mass, radius and  T$_{\rm eff}$ for both stars in this binary system. 
We describe a technique we have developed to measure the flux ratio at near-infrared wavelengths from the NIRPS spectra.
This substantially improves the accuracy of our T$_{\rm eff}$ measurement for the M-dwarf in this binary.

In Section~\ref{sec:obs} we describe the observations we have used for our analysis.
In Section \ref{Analysis} we use light curve data along with HARPS and NIRPS spectra to determine the radial velocities, rotational broadening, masses, radii, and effective temperature of both components of CD$-$27~2812.
We use the HARPS spectra to determine the metallicity of the primary star, and the NIRPS spectra to determine the flux ratios in the J and H bands.   
In Section \ref{sec:discuss} we discuss the implications of our results and we give our conclusions in Section \ref{sec:conclusions}.

\section{Observations}
\label{sec:obs}
\subsection{TESS photometry}
CD$-$27 2812 was observed at 120-s cadence by TESS \citep{2015JATIS...1a4003R} in 2 sectors in 2018 and 2024, covering four primary and six secondary eclipses. We used the \software{pdc\_sapflux} produced by the TESS Science processing Operation Center (SPOC). The light curve data was downloaded from MAST using \software{Lightkurve}\citep{2018ascl.soft12013L}\footnote{\url{https://docs.lightkurve.org/}} using the default quality flag bitmask to mask out bad cadences. Using the values from Table 1 of \cite{2023MNRAS.522.2683M} we were able to divide the light curve into sections of complete eclipses plus data on either side. 
For each section of the light curve, we divided the data by a straight line fit by least squares to the data either side of the eclipse to remove trends in the data due to instrumental noise and intrinsic stellar variability.

\subsection{HARPS + NIRPS spectroscopy}
We observed CD-27~2812 on 9 nights from 2023 October 4 to 2024 March 30 using the HARPS  and NIRPS  spectrographs on the ESO (European Southern Observatory) 3.6-m telescope.
The HARPS spectra cover the wavelength range 378.3\,--\,691.4\,nm at a resolving power $R\approx 80\,000$. 
We obtained two spectra per night with HARPS, each with a typical signal to noise ratio SNR~$\approx 70$.
We used the 1-dimensional spectra produced automatically by the HARPS data reduction pipeline DRS~3.8 by the observatory.
The NIRPS spectra cover the wavelength range 966\,--\,1923\,nm at a resolving power $R\approx 70\,000$. 
We obtained three spectra per night with NIRPS, each with a typical signal to noise ratio SNR~$\approx 80$.
We used the data extracted order-by-order produced from the raw data using NIRPS pipeline version 7.3.2.

We measured the radial velocity (RV) of the primary star by performing a cross correlation between the HARPS spectra and the numerical mask HARPS\_SOPHIE.G2.375\_679nm provided with the software package \software{iSpec} \citep{2019MNRAS.486.2075B, 2014A&A...569A.111B}.
We also measured the primary RV in the NIRPS spectra using the NIRPS G2 spectral mask provided with the NIRPS data reduction pipeline. 
These RV measurements are provided in Table~\ref{tab:RV}.

\section{Analysis}
\label{Analysis}

\subsection{TESS photometry and primary star radial velocity fit} 
\label{sec:jktebop_TESS}
We used the Nelson-Davis-Etzel (NDE) light curve model \citep{1972ApJ...174..617N}  implemented in the software package  \software{jktebop}\footnote{\url{http://www.astro.keele.ac.uk/jkt/codes/jktebop.html}} \citep{2010MNRAS.408.1689S} to analyse the TESS light curve together with the radial velocity measurements of the primary star.

The ellipsoidal effect, gravity darkening and the reflection effect were ignored in the analysis of the light curves because the stars are well separated, i.e. very nearly spherical. 
Limb darkening was modelled using the power-2 law. 
The values of the parameters $h_1$ and $h_2$ were estimated by interpolation within the relevant table from \citet{2018A&A...616A..39M}.
The effect of the assumed value of $h_1$ for the primary star can be seen in the curvature of the light curve at the bottom of the primary eclipse so this parameter was allowed to vary in the analysis of the light curve. 
The effect of $h_2$ for the primary star on the light curve model is very subtle so this parameter was fixed at the value obtained from \citet{2018A&A...616A..39M}. 
The assumed limb darkening of the secondary star has a negligible effect on the light curve so we used the fixed nominal values $h_1=0.8$ and $h_2=0.4$.

Standard errors on RVs (given in Table~\ref{tab:RV}) and light curve data  were set equal to root mean square of the residuals (rms) from preliminary fits to each data set separately to ensure correct relative weighting of the different data sources. 

The value for the flux from the target in the photometric aperture provided in the meta data for the SPOC light curves is $\mbox{\sc crowdsap} = 0.9770$. 
This is accounted for in the PDCSAP light curves but there will be some uncertainty in this correction. 
To account for this uncertainty, which we assume to be 1\,per~cent,  we assume $\ell_3 = 0.00 \pm 0.01$ as a prior in the least-squares fit.

The details of the free parameters of the least-squares fit are similar to those described in Section 2.1 of \citet{2024MNRAS.531.4577M} except that we also include following free parameters for the primary star spectroscopic orbit: the semi-amplitude, $K_1$; the radial velocity of the binary barycentre measured using HARPS data, $V_{0{\rm HARPS}}$; the offset between the HARPS and NIRPS RV measurements, $V_{0,{\rm NIRPS}} - V_{0,{\rm HARPS}}$. 
The work-around needed to include the latter parameter in the least-squares fit using {\sc jktebop} is described in Appendix~\ref{sec:jktebop}.

\begin{figure}
    \centering
    \includegraphics[width=1\linewidth]{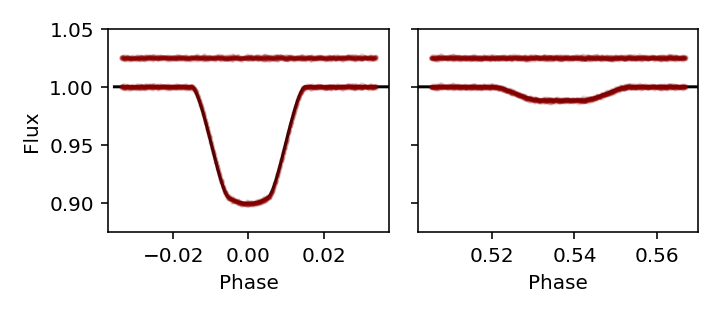}
    \caption{TESS photometry of CD$-$27 2812 as a function of orbital phase (red points) and best-fit light curve models for the data fitted (black lines). The residuals from the best-fit models are shown offset vertically above the light curve data.}
    \label{fig:lcfit}
\end{figure}

\begin{figure}
    \centering
    \includegraphics[width=1\linewidth]{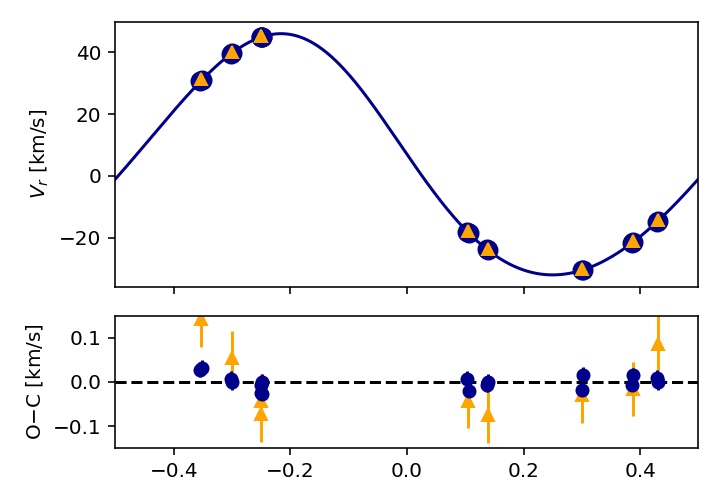}
    \caption{Radial velocity curve for the primary component of CD$-$27~2812 using the HARPS (circles) and NIRPS (triangles) data, with residuals in the lower panel. The NIRPS RV data has been corrected for the zero point offset from the HARPS data.}

    \label{fig:RV1 curve}
\end{figure}

The best fit to the TESS light curve in shown in Fig.~\ref{fig:lcfit} and the best fit to the primary star RV measurements is shown in Fig.~\ref{fig:RV1 curve}. 
The best fit model parameters and their standard errors estimates are given in Table~\ref{tab:lcfit}. 
The uncertainties were calculated using 10\,000 iterations of a Monte Carlo simulation.

\begin{table*}
\caption[]{Mean and standard error of the mean for free parameters and derived quantities from least-squares fits to the TESS light curve of CD$-$27~2812. }
\label{tab:lcfit}
\begin{center}
  \begin{tabular}{@{}lrl}
\hline
\noalign{\smallskip}
 \multicolumn{1}{@{}l}{Parameter} &
 \multicolumn{1}{l}{Value} &
 Note \\
\noalign{\smallskip}
\hline
\noalign{\smallskip}
\multicolumn{3}{@{}l}{Model parameters} \\
\noalign{\smallskip}
BJD$_{\rm TDB}$ T$_0$ & $2458469.611103 \pm 0.000039 $ & Time of mid-transit \\
$P$    & $ 7.8357417365 \pm 0.00000018 $ & Orbital period in days \\
$ r_1+r_2      $&$      0.10907\pm     0.00015 $& Sum of the radii\\
$ k=r_2/r_1    $&$     0.3084 \pm       0.0015 $& \\
$ h_1          $&$      0.8133 \pm       0.0041 $& Primary star limb-darkening parameter\\
$ i            $&$     87\fdg3 \pm      0\fdg02 $& Orbital inclination \\
$ \ell_3       $&$    0.0023 \pm      0.0095 $& Third light. Constrained by a prior.\\
$ e\cos(\omega)$&$   0.05757 \pm    0.00004 $& \\
$ e\sin(\omega)$&$  0.05337 \pm  0.00015 $&  \\
$ \ell_T       $&$      0.01178 \pm      0.00012 $& Flux ratio in TESS band\\
\noalign{\smallskip}
\multicolumn{3}{@{}l}{Derived parameters} \\
\noalign{\smallskip}
$ e            $&$    0.0785 \pm    0.0001 $& Orbital eccentricity\\
$ J            $&$       0.1238 \pm       0.0004$&Average surface brightness ratio $ =\ell_T/k^2$\\
$ r_1          $&$      0.0834 \pm      0.0002 $& Fractional stellar radius, star 1\\
$ r_2          $&$      0.02571 \pm      0.00007 $& Fractional stellar radius, star 2 \\
$K_1          $&$  39.073   \pm 0.006 $& Primary radial velocity amplitude [km/s]\\
$V_{0,{\rm HARPS}} $&$   4.7327    \pm 0.0045  $ &  Systemic radial velocity, HARPS data [km/s] \\
$V_{0,{\rm NIRPS}} - V_{0,{\rm HARPS}} $&$   0.19    \pm 0.02 $ &  Zero-point offset   [km/s] -- see Appendix \ref{sec:jktebop}\\
\noalign{\smallskip}
\multicolumn{3}{@{}l}{Goodness of fit} \\
\noalign{\smallskip}
rms (HARPS)    &  17 m/s    & root mean square residual for HARPS data \\
rms (NIRPS)   &  56 m/s    & root mean square residual for NIRPS data \\
rms (TESS)     &  0.8 mmag  & root mean square residual for TESS data \\
\noalign{\smallskip}
\hline
\end{tabular}
\end{center}
\end{table*}


\begin{table}
    \centering
    \begin{tabular}{rrrl}
    \hline
\multicolumn{1}{l}{Time[BJD]} & 
\multicolumn{1}{l}{Velocity [km/s]} & 
\multicolumn{1}{l}{Error} &
\multicolumn{1}{l}{Source} \\
    \hline
2460222.856600 &       44.930 &      0.016 & HARPS \\
2460222.863596 &       45.099 &      0.056 & NIRPS \\
2460222.869500 &       45.040 &      0.016 & HARPS \\
2460300.812600 &       39.560 &      0.016 & HARPS \\
2460300.821661 &       39.931 &      0.056 & NIRPS \\
2460300.828600 &       39.840 &      0.016 & HARPS \\
2460366.674600 &      $-$18.100 &      0.016 & HARPS \\
2460366.683319 &      $-$18.125 &      0.056 & NIRPS \\
2460366.690300 &      $-$18.510 &      0.016 & HARPS \\
2460379.570000 &       44.900 &      0.016 & HARPS \\
2460379.579169 &       45.135 &      0.056 & NIRPS \\
2460379.586400 &       45.030 &      0.016 & HARPS \\
2460382.609600 &      $-$23.840 &      0.016 & HARPS \\
2460382.618475 &      $-$23.850 &      0.056 & NIRPS \\
2460382.625400 &      $-$24.130 &      0.016 & HARPS \\
2460384.562000 &      $-$21.690 &      0.016 & HARPS \\
2460384.570786 &      $-$21.377 &      0.056 & NIRPS \\
2460384.578200 &      $-$21.370 &      0.016 & HARPS \\
2460386.592900 &       30.590 &      0.016 & HARPS \\
2460386.601587 &       31.068 &      0.056 & NIRPS \\
2460386.608900 &       30.990 &      0.016 & HARPS \\
2460399.556100 &      $-$30.600 &      0.016 & HARPS \\
2460399.565322 &      $-$30.362 &      0.056 & NIRPS \\
2460399.572700 &      $-$30.440 &      0.016 & HARPS \\
2460400.566100 &      $-$14.900 &      0.016 & HARPS \\
2460400.574761 &      $-$14.475 &      0.056 & NIRPS \\
2460400.582100 &      $-$14.550 &      0.016 & HARPS \\

       \noalign{\smallskip}
    \hline
    \end{tabular}
    \caption{Radial velocities from HARPS and NIRPS CCF using HARPS SOPHIE G2 Mask and NIRPS G2 Mask.}
    \label{tab:RV}
\end{table}

\subsection{Rotational broadening and metallicity measurements}
We determined the rotational broadening of the primary star($v_1\sin i$) by generating synthetic spectra in {\sc iSpec}, as described in \citep{2023MNRAS.522.2683M} using the default settings, with varying $v_1\sin i$ until the width of the cross-correlation function matched that measured using the HARPS spectra.
We determined the $v_1 \sin i$ to be $6.95 \pm 0.1$\,km/s where the error is estimated by accounting for the uncertainty in the calibration of the macroturbulence velocity, $v_{\rm macro}$, from \citet{2010MNRAS.405.1907B}.

\begin{figure}
    \centering
    \includegraphics[width=1\linewidth]{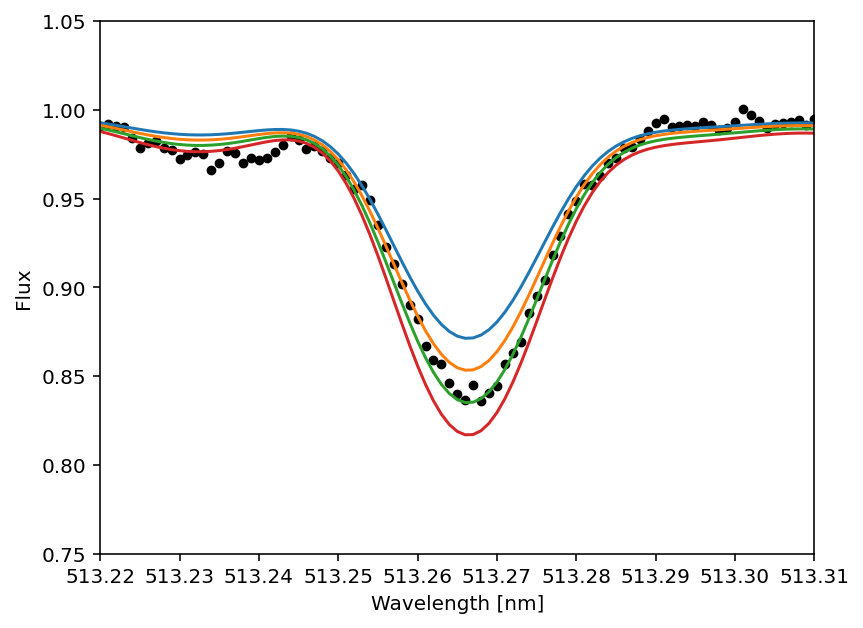}
    \caption{The median HARPS spectrum (dots) of the primary star and synthesised spectra with metallicities of -0.1 to 0.2 (top to bottom, in steps of 0.1). The HARPS spectrum closely matches the [Fe/H] = 0.1 spectra at the 513.27 nm Fe II spectra line.}
    \label{fig:metal}
\end{figure}

Using this measurement we were then able to create synthetic spectra over a grid of metallicities so that we can estimate [Fe/H] for the primary star. 
We visually compared these synthetic spectra to the median HARPS spectrum  in the region of 58 Fe~I and Fe~II lines at wavelengths selected from \citet{2013MNRAS.428.3164D} (their Table~2).  
This median HARPS spectrum has has had the signal from the M-dwarf removed using a similar method to that described for the NIRPS spectra in Section \ref{sec:NIRPS bands}.
An example for one Fe~I line is shown in Fig.~\ref{fig:metal}. 
This  high signal-to-noise  spectrum of the primary star (S/N $\approx 700$ at 500\,nm) is available from the corresponding author's website\footnote{\url{https://pmaxted.github.io/BenchmarkDEBS/}} and in the supplementary online information that accompanies this article.
The mean of the best fitting metallicities is [Fe/H] = 0.15. We assign this estimate an uncertainty of 0.15, based on the discussion in \cite{2019ARA&A..57..571J}.

\subsection{Detection of the M dwarf in the NIRPS spectra}
\label{sec:ccfstack}
To measure the semi-amplitude of M~dwarf's spectroscopic orbit, $K_2$, we used the method described in \cite{2022MNRAS.513.6042M}. 
This method is based on calculating the average cross-correlation function (CCF) against a suitable template spectrum after removing the contribution of the primary star from the observed spectra and then shifting these CCFs to the rest frame of the M~dwarf assuming a range of $K_2$ values.
The barycentric radial velocity of the M~dwarf at the time of mid-exposure for each spectrum is
 \begin{equation}
 \label{eq:K2}
 V_{r,2} = K_2\,[\cos(\nu+\omega_2)+e\cos(\omega_2)]
 \end{equation}
 The value of the eccentricity $e$ and the longitude of periastron $\omega_2 = \omega_1+\pi$ are known accurately from the spectroscopic orbit of the primary star. Similarly, the true anomaly at the time of mid-exposure, $\nu$, can be accurately predicted from the values of $T_0$ (time of mid-transit), $P$ (orbital period),  $e$ and $\omega_1$ obtained from the light curve fitting.

We use synthetic spectra calculated with BT-NextGen model atmospheres taken from \citet{BT-NG} as templates for the spectra of the M dwarf, using linear interpolation to create a spectrum appropriate for $T_{\rm eff}=3600$\,K, ${\rm [Fe/H]} = 0.0$, $\log g =  4.75$ and $[\alpha/{\rm Fe}] = 0.0$.
The cross-correlation function is calculated order-by-order. 
Low-frequency noise in the data for each order was removed prior to cross-correlation using a 5$^{\rm th}$ order high-pass Butterworth filter with a critical frequency of 32/4096 pixels$^{-1}$. 
The data are apodized using a Gaussian filter with a standard deviation of 64 pixels applied to the data at each end of the order. 
The correlation coefficient for each order is calculated after shifting the template according to radial velocity computed with equation (\ref{eq:K2}) using weights calculated from the estimated standard errors on each pixel.
This is repeated for a uniform grid of $K_2$ values. 
The average CCF over all orders in the H~band as a function of $K_2$ (``stacked CCF'') is shown in Fig.~\ref{fig:CCF_figure}. 
We expect $K_2\approx 92$\,km\,s$^{-1}$ based on the initial mass ratio estimate from \citet{2023MNRAS.522.2683M}.
There is indeed a peak in the stacked CCF near this value of $K_2$. 
To measure the position of this peak we model the stacked CCF as a Gaussian process (GP) plus a Gaussian profile. 
We use the \software{celerite} package \citep{celerite1} to compute the likelihood for a GP with a kernel of the form $k(\tau) = a_j\,e^{-c_j\,\tau}$ and the affine-invariant Markov chain Monte Carlo sampler \software{emcee} \citep{2013PASP..125..306F} to sample the posterior probability distribution for the model parameters. 
The results for the Y, J and H bands are given in Table~\ref{tab:K2} together with the weighted average values computed using the {\sc combine} algorithm described in Appendix~A of \citet{2022MNRAS.514...77M}. 

\begin{figure}
    \centering
    \includegraphics[width=1\linewidth]{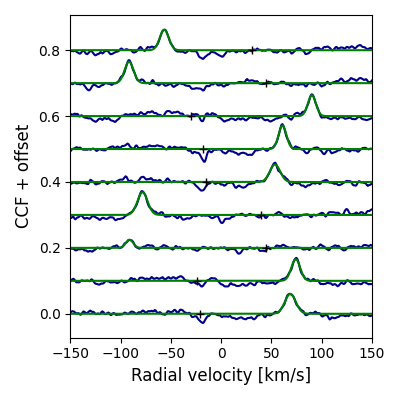}
    \caption{Cross-correlation functions of the individual NIRPS spectra against an M-dwarf template after removal of the contribution from the primary star.
    A Gaussian profile fit to the peak due to the M-dwarf is shown for each CCF. 
    The cross marks the radial velocity of the primary star at the date of observation for each spectrum.
    \label{fig:ccfs}}
\end{figure}

Following advice from an anonymous referee, we inspected the CCFs from the individual spectra to better understand the reliability of our M-dwarf detection.
The results can be seen in Fig.~\ref{fig:ccfs}.
It can be seen that the M-dwarf is clearly detected in the individual spectra and that the primary star contribution has been accurately removed.

\begin{table}
\caption[]{Summary of analysis for stacked CCFs from NIRPS spectra in the Y, J and H bands.}
\label{tab:K2}
\begin{center}
  \begin{tabular}{lrrr}
\hline
\noalign{\smallskip}
 \multicolumn{1}{l}{Band} &
 \multicolumn{1}{l}{$K_2$ [km/s]} &
 \multicolumn{1}{l}{Correction Factor} &
 \multicolumn{1}{l}{Wavelength Range [nm]}\\
\noalign{\smallskip}
\hline
\noalign{\smallskip}
Y & $ 94.59 \pm 0.13 $ & 0.717 &  980 - 1100\\
J & $ 94.03 \pm 0.25 $ & 1.022 & 1211 - 1298\\
H & $ 94.41 \pm 0.07 $ & 0.987 & 1517 - 1709\\
\noalign{\smallskip}
\hline
Average & $ 94.42 \pm 0.08 $ &   \\
\noalign{\smallskip}
\hline

\end{tabular}
\end{center}
\end{table}

We calculate another set of cross-correlation functions in the H-band using synthetic spectra to determine the value of $v_2\sin i$, the rotational broadening of the companion star. 
The synthetic spectra are copies of the template spectrum used to compute the stacked CCF described above, but with rotational broadening applied and shifted to the radial velocity of the M~dwarf observed on each night.
The assumed value of $v_2\sin i$ was adjusted until the 0.33, 0.5, and 0.66 points on the synthetic and observed stacked CCFs matched, as can be seen in Fig.~\ref{fig:Template-Template}. 
We estimate  $v_2\sin i = 3.0\pm0.2$\,km\,s$^{-1}$ using this method. 
We also verified that the value of $K_2$ measured from the synthetic stacked CCF is consistent with the input value to within the quoted error.

\subsection{Flux ratios in the Y, J, and H bands}
\label{sec:NIRPS bands}
We attempted to measure the flux ratio in the Y, J and H bands from the height of the peak in the stacked CCFs.
We did this by implementing a 'flux correction factor' ($CF$) that could be applied to the model spectra and adjusted such that subtracting the scaled model spectrum from the observed spectra removes the signal of the M dwarf from the stacked CCF.

The scaling factor by which flux of the model spectra is multiplied by is calculated with: 
\begin{equation}
 S =  CF \times L_{\rm TESS} \times \frac{\int(F_{1}\times R_{\rm TESS}(\lambda)\times \lambda)}{\int(F_{2}\times R_{\rm TESS}(\lambda)\times \lambda)}
\end{equation}
where $L_{\rm TESS}$ is the flux ratio in the TESS bands, $F_{1,2}$ are the fluxes of the primary and secondary model spectra, and $R_{TESS}$ is the interpolation of the {\rm TESS} response function over the wavelength of the model spectra.  

We generated a range of synthetic template spectra with differing values of $CF$ and interpolated to find the value of $CF$ that removed the M-dwarf signal from the CCF in each band. 
The height of the CCF peak was measured using a least-squares Gaussian fit with the width and location of the Gaussian profile fixed to the values measured from the stacked CCF before subtracting the M dwarf signal.
We can then find the value of $CF$ that yields a peak height of 0 in the CCF,  as shown in Fig. \ref{fig:CF_interp}. 
We expect the $CF$ values to be around 1 for each band, but not exactly 1 because of inaccuracies in the model spectra. 
The $CF$ values we obtained can be seen in Table \ref{tab:K2}, however, the $CF$ for the Y band was significantly lower than expected.  
Whilst the peak in the Y band was clear enough to produce a $K_2$ value consistent with the J and H bands, the signal in the CCF was too noisy to provide a reliable flux ratio measurement, and thus we have dropped the Y band in further analysis. 
 
\begin{figure}
	\includegraphics[width=\columnwidth]{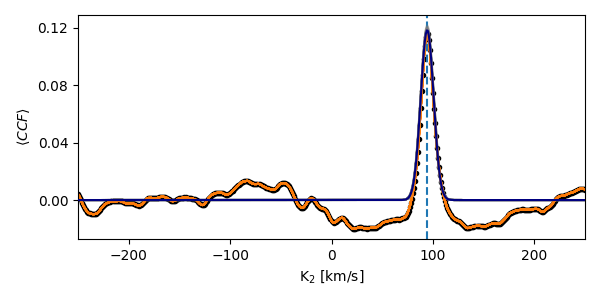}
	\includegraphics[width=\columnwidth]{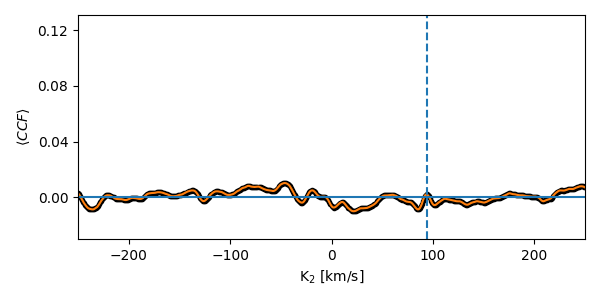}
    \caption{{\it Upper panel:}  Mean cross-correlation function of CD$-$27~2812 using H-band spectra after shifting to the rest frame of CD$-$27~2812\,B assuming a range of $K_2$ values. The vertical dashed line marks $K_2=94$\,km/s. Gaussian process fit (orange) of a Gaussian profile to the peak near $K_2=94$\,km/s in the stacked CCF (black points). The maximum-likelihood Gaussian profile is plotted in dark blue and 50 samples from the posterior probability distribution are plotted in light grey. 
    {\it Lower panel:} Same as the upped panel but applied to H-band spectra after dividing by a model M-dwarf spectrum.
    }
    \label{fig:CCF_figure}
\end{figure}

\begin{figure}
    \includegraphics[width=\columnwidth]{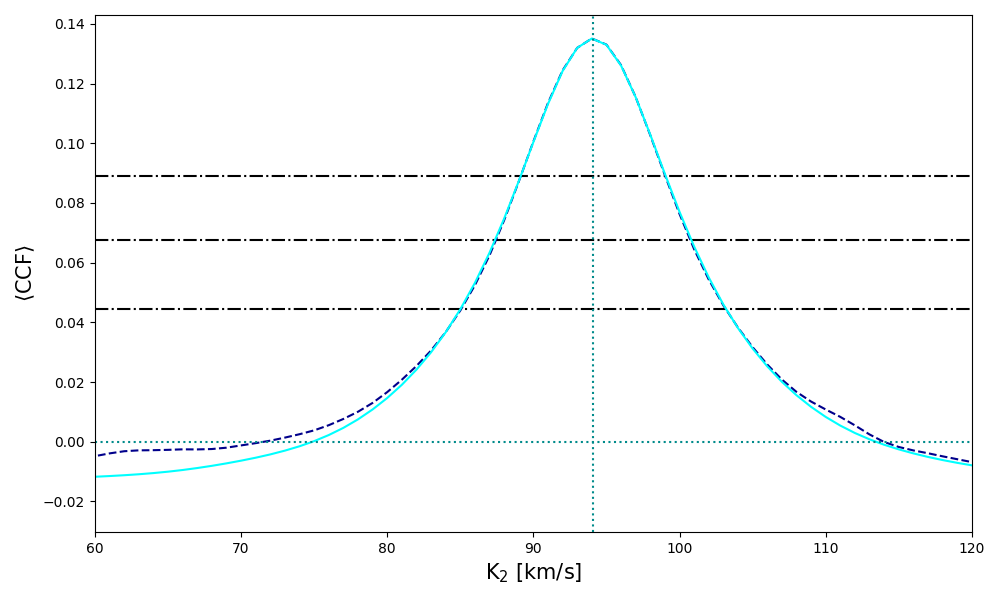}
    \caption {Cross-correlation functions in the H band of the mean spectra and a model with no broadening (dashed line), and the model against a broadened model (solid line), where the two models match at the 2/3 point. Dashed-dot lines are the 0.33, 0.5, and 0.66 points of the mean spectra CCF. 
    }

    \label{fig:Template-Template}
\end{figure}

\begin{figure}
	\includegraphics[width=\columnwidth]{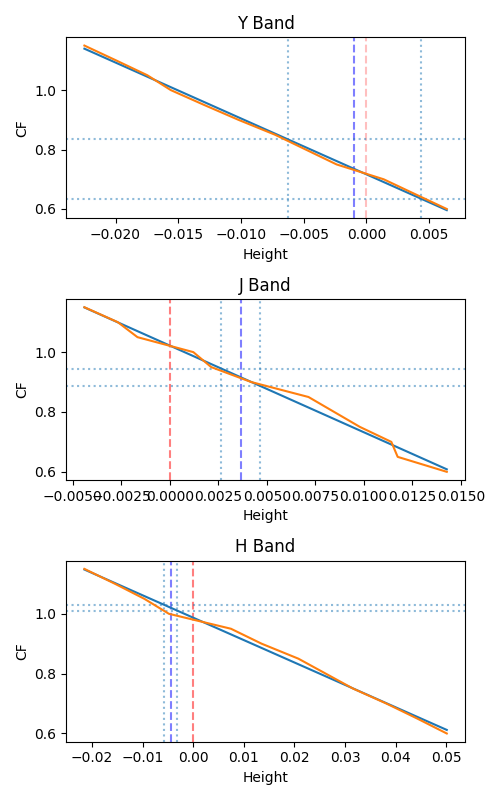}
    \caption{Cross correlation amplitude peak against correction factor (orange line) for each band. A linear fit (blue line) was used to determine the correction factor that would produce a peak height of zero (red dashed line). The resulting peak height (blue dashed line) and its errors (blue dotted lines) are also plotted.}
    \label{fig:CF_interp}
\end{figure}

We obtained the flux ratio in the J band by using:
\begin{equation}
   \ell_{\rm J} =\frac{\int (S\,F_{2}\times R_{\rm NIRPS\_J}(\lambda)\times \lambda)}{\int(F_{1}\times R_{\rm NIRPS\_J}(\lambda)\times \lambda)}
   \label{eq:factorflx}
\end{equation}
 where $R_{\rm NIRPS\_J}$ is the interpolation of the NIRPS response function obtained from version P116 1.0.5 of the NIRPS exposure time calculator \citep{2020SPIE11449E..1BB}.\footnote {\url{https://etc.eso.org/nirps} assuming airmass 1.2} This was repeated for the H band to obtain $\ell_{\rm H}$. 
To estimate the standard error on these flux ratios, we took the rms  of the residuals in the stacked CCF after removing the M dwarf signal (Fig.~\ref{fig:CCF_figure}, lower panel) and propagated this error through the height - CF relations shown in Fig.~\ref{fig:CF_interp}.
The final column of Table \ref{tab:mags} shows our final flux ratios.

\subsection{Mass and radius}
\label{sec:mr}
The mass, radius of the two stars with their standard errors have been computed in nominal solar units \citep{2016AJ....152...41P} using a Monte Carlo simulation.
The random sample of 10\,000 values for $r_1=R_1/a$, $r_2=R_2/a$, $i$, $e$ and $K_1$ generated using {\sc jktebop} (Section \ref{sec:jktebop_TESS}) were paired with a random Gaussian sample of $K_2 = 94.42\pm0.08$\,km/s values.     
The error on $P=7.8357349130$\,d is assumed to be negligible. 
The results are given in Table~\ref{tab:mr}

\subsection{Effective temperature measurements}
\label{sec:teb}

f We used the software package {\sc teb} \citep{2020MNRAS.497.2899M} to measure the effective temperature of the stars in CD$-$27~2812. The effective temperature (T$_{\rm eff}$) of a star is defined by the equation  
\[L=4\pi R^2 \sigma_{\rm SB} {\rm T}_{\rm eff}^4,\]
where $R$ is the Rosseland radius, $L$ in the luminosity and  $\sigma_{\rm SB}$ is the Stefan-Boltzmann constant. 
For a binary star with parallax $\varpi$, i.e. at a distance $d=1/\varpi$, the bolometric flux corrected for extinction observed at the top of Earth's atmosphere is
\[f_{0,b}= f_{0,1}+f_{0,2}=\frac{\sigma_{\rm SB}}{4}\left[\theta_1^2{\rm T}_{\rm eff,1}^4 + \theta_2^2{\rm T}_{\rm eff,2}^4\right],\]
where $\theta_1=2R_1\varpi$ is the angular diameter of star 1, and similarly for star 2 , and $f_{0,1}$ and $f_{0,2}$ are the extinction-corrected bolometric fluxes of the two stars. All the quantities are known or can be measured for CD$-$27 2812 provided we can accurately integrate the observed flux distributions for the two stars independently. 

The M-dwarf contributes less than 1.5\,per~cent to the total flux so it is not necessary to make a very accurate estimate of the M-dwarf flux distribution in order to derive an accurate value of $T_{\rm eff}$ for the F9\,V primary star.
With the flux ratio measurements in the J and H bands from the NIRPS spectroscopy in addition to the flux ratio measurement in the TESS band, we can also make an accurate measurement of T$_{\rm eff}$ for the M-dwarf companion.

The photometry used in this analysis is given in Table~\ref{tab:mags}. 
The Gaia photometry is from Gaia data release DR3. J, H and K$_{s}$ magnitudes are from the 2MASS survey \citep{2006AJ....131.1163S}. 
WISE W1, W2 and W3 magnitudes \citep{2010AJ....140.1868W,2011ApJ...731...53M} are from the All-Sky Release Catalog.\footnote{\url{https://wise2.ipac.caltech.edu/docs/release/allwise/}}  
B$_{\rm T}$ and V$_{\rm T}$ magnitudes are taken from the Tycho-2 catalogue \citep{2000A&A...355L..27H}.
The u- and v-band photometry are taken from the SkyMapper survey DR4 \citep{2024PASA...41...61O}.
We were unable to use the NUV magnitude from GALEX as there were no reliable measurements available in the GALEX archive.


\begin{table*}
\caption{Observed magnitudes, colours and flux ratios for CD-27~2812 and predicted values based on our synthetic photometry. The
predicted magnitudes are shown with error estimates from the uncertainty on the zero-points for each photometric system.  The
pivot wavelength for each band pass is shown in the column headed  $\lambda_{\rm pivot}$.  The estimated apparent magnitudes for
each star are shown given in the columns headed $m_1$ and headed $m_2$.  The flux ratio in each band is shown in the final
column. }
\label{tab:mags}
\centering
\begin{tabular}{@{}lrrrrrrr}
\hline
Band &  $\lambda_{\rm pivot}$ [nm]& \multicolumn{1}{c}{Observed} &\multicolumn{1}{c}{Computed} &
\multicolumn{1}{c}{$\rm O-\rm C$} &\multicolumn{1}{c}{$m_1$}  & \multicolumn{1}{c}{$m_2$}  &
\multicolumn{1}{c}{$\ell$} [\%] \\
\hline
\noalign{\smallskip}
u   &   349.3 & $11.183\pm 0.011 $& $11.160\pm 0.030 $& $+0.023 \pm 0.032 $& $11.160\pm 0.030 $ & $19.845\pm 0.030 $ &  0.03 \\
v   &   383.6 & $10.770\pm 0.013 $& $10.790\pm 0.020 $& $-0.020 \pm 0.024 $& $10.790\pm 0.020 $ & $19.028\pm 0.020 $ &  0.05 \\
B$_{\rm T}$  &   421.2 & $10.391\pm 0.028 $& $10.380\pm 0.014 $& $+0.011 \pm 0.031 $& $10.382\pm 0.014 $ & $17.759\pm 0.014 $ &  0.11 \\
G$_{\rm BP}$ &   511.0 & $ 9.893\pm 0.003 $& $ 9.894\pm 0.006 $& $-0.001 \pm 0.007 $& $ 9.897\pm 0.006 $ & $16.168\pm 0.006 $ &  0.31 \\
V$_ {\rm T}$  &   533.5 & $ 9.834\pm 0.024 $& $ 9.808\pm 0.014 $& $+0.026 \pm 0.028 $& $ 9.812\pm 0.014 $ & $16.101\pm 0.014 $ &  0.31 \\
G   &   621.8 & $ 9.631\pm 0.003 $& $ 9.631\pm 0.008 $& $-0.001 \pm 0.008 $& $ 9.638\pm 0.008 $ & $15.124\pm 0.008 $ &  0.64 \\
G$_{ \rm RP}$ &   776.9 & $ 9.207\pm 0.004 $& $ 9.201\pm 0.004 $& $+0.006 \pm 0.006 $& $ 9.213\pm 0.004 $ & $14.111\pm 0.004 $ &  1.10 \\
J   &  1240.6 & $ 8.765\pm 0.027 $& $ 8.743\pm 0.015 $& $+0.022 \pm 0.031 $& $ 8.770\pm 0.015 $ & $12.805\pm 0.015 $ &  2.43 \\
H   &  1649.0 & $ 8.520\pm 0.055 $& $ 8.474\pm 0.019 $& $+0.046 \pm 0.058 $& $ 8.511\pm 0.019 $ & $12.169\pm 0.019 $ &  3.44 \\
K$_s$  &  2162.9 & $ 8.493\pm 0.031 $& $ 8.405\pm 0.030 $& $+0.088 \pm 0.043 $& $ 8.447\pm 0.030 $ & $11.962\pm 0.030 $ &  3.92 \\
W1  &  3389.7 & $ 8.403\pm 0.023 $& $ 8.389\pm 0.036 $& $+0.014 \pm 0.043 $& $ 8.435\pm 0.036 $ & $11.851\pm 0.036 $ &  4.30 \\
W2  &  4640.6 & $ 8.437\pm 0.020 $& $ 8.423\pm 0.059 $& $+0.014 \pm 0.062 $& $ 8.471\pm 0.059 $ & $11.822\pm 0.059 $ &  4.57 \\
W3  & 12567.5 & $ 8.419\pm 0.023 $& $ 8.402\pm 0.053 $& $+0.017 \pm 0.058 $& $ 8.452\pm 0.053 $ & $11.761\pm 0.053 $ &  4.75 \\
\noalign{\smallskip}
\multicolumn{5}{@{}l}{Flux ratios [\%]} \\
\noalign{\smallskip}
TESS &   788.0 & $ 1.178 \pm 0.012 $& $ 1.194 $& $-0.016 \pm 0.012 $ \\
Nirps\_J &  1253.0 & $ 2.700 \pm 0.300 $& $ 2.500 $& $+0.200 \pm 0.300 $ \\
Nirps\_H &  1605.4 & $ 3.610 \pm 0.080 $& $ 3.394 $& $+0.216 \pm 0.080 $ \\
\noalign{\smallskip}
\hline
\end{tabular}
\end{table*}

To estimate the reddening towards CD$-$27~2812 we use the calibration of E(B$-$V) versus the equivalent width of the interstellar Na\,I~D lines by \citet{2025RNAAS...9..146M}. 
To measure the equivalent widths of the interstellar Na\,I~D lines from the HARPS spectra, we used the method described in \citet{2025MNRAS.544.4611M}. 
The equivalent width of the Na\,I~D$_2$ line measured by numerical integration is EW(Na\,I~D$_2) = 16 \pm 1$\,m\AA\ which implies E(B$-$V$)= 0.006 \pm 0.014$.
The result from the Na\,I~D$_1$ line is consistent with this estimate.
We use this as a Gaussian prior in our analysis but exclude negative values of E(B$-$V).

The method we have used to measure $T_{\rm eff}$ for eclipsing binary stars is described fully in \citet{2020MNRAS.497.2899M}. The results of CD-27~2812 presented here were computed using version 2025.10.15 of the software. This version is similar to the updated version of the software described in \citet{2025MNRAS.544.4611M}. 
This new version of the software changes the way that synthetic photometry is calculated by the convolution of the instrument response function with the synthetic spectrum. 
The spectrum is now interpolated onto the same wavelength sampling as the response function, rather than the other way around.
This enables us to better control the sampling used for the calculation of the synthetic photometry.
The NIRPS response functions obtained from the NIRPS pipeline software were binned into 10\,nm bins, providing a smoother function and increasing the speed of the calculation. 

The {\sc teb} software uses \software{emcee} \citep{2013PASP..125..306F} to sample the posterior probability distribution (PPD) $P(\Theta| D)\propto P(D|\Theta)P(\Theta)$ for the model parameters $\Theta$ with prior $P(\Theta)$ given the data, $D$ (observed apparent magnitudes and flux ratios). 
The model parameters are  $$\Theta = \left({\rm T}_{\rm eff,1},  {\rm T}_{\rm eff,2}, \theta_1, \theta_2, {\rm E}({\rm B}-{\rm V}), \sigma_{\rm ext}, \sigma_{\ell},  c_{1,1}, \dots, c_{2,1}, \dots\right).$$  
The prior $P(\Theta)$ is calculated using the angular diameters $\theta_1$ and $\theta_2$ derived from the radii $R_1$ and $R_2$ and the parallax $\varpi$, the priors on the flux ratio at infrared wavelengths based on the colour\,--\,T$_{\rm eff}$ relations, and the Gaussian prior on the reddening described above. 
The hyper-parameters $\sigma_{\rm m}$ and $\sigma_{\ell}$ account for additional uncertainties in the synthetic magnitudes  and flux ratio, respectively, due to errors in zero-points, inaccurate response functions, stellar variability, etc. 
The parallax ($\varpi = 5.158\pm0.020$ mas) is taken from Gaia EDR3 with corrections to the zero-point from \citet{2022MNRAS.509.4276F}.

To calculate the synthetic photometry for a given value of $T_{\rm eff}$ we used a model spectral energy distribution (SED) multiplied by a distortion function, $\Delta(\lambda)$. The distortion function is a linear superposition of Legendre polynomials in log wavelength. The coefficients of the distortion function for star 1 are $c_{1,1}, c_{1,2}, \dots$, and similarly for star 2. The distorted SED for each star is normalized so that the total apparent flux prior to applying reddening is $\sigma_{\rm SB}\theta^2{\rm T}_{\rm eff}^4/4$. These distorted SEDs provide a convenient function that we can integrate to calculate synthetic photometry that has realistic stellar absorption features, and where the overall shape can be adjusted to match the observed magnitudes from ultraviolet to infrared wavelengths. This means that the effective temperatures we derive are based on the integrated stellar flux and the star's angular diameter, not SED fitting.

For this analysis we use model SEDs interpolated from the grid of synthetic spectra computed from BT-Settl model atmospheres \citep{2013MSAIS..24..128A} obtained from the Spanish Virtual Observatory.\footnote{\url{http://svo2.cab.inta-csic.es/theory/newov2/index.php?models=bt-settl}} 
For the F9\,V star appropriate we use the model with  $T_{\rm eff}=6150$\,K, $\log g \approx 4.0$, $[{\rm Fe/H}] = 0.04$ and   $[{\rm \alpha/Fe}] = 0.0$. 
For the reference SED for the M dwarf companion we assume $T_{\rm eff}=3600$\,K, $\log g = 4.75$, and the same composition. 
We experimented with distortion functions with 2, 3, 4 and 5 coefficients per star and found the results to be very similar in all cases. 
The results presented here use four distortion coefficients per star because this number of coefficients resulted in the minimum value of the Bayesian information criterion (BIC) in a set of initial runs.
For the results here we used 300 walkers and a burn-in of 10\,000 steps in {\sc emcee} before running the sampler for 600 steps for the calculation of the results given here. 
Convergence of the sampler was checked by visual inspection of the parameter values for each walker as a function of step number.
The predicted apparent magnitudes including their uncertainties from errors in the zero-points for each photometric system are compared to the observed apparent magnitudes in Table~\ref{tab:mags}.

\begin{figure}
	\includegraphics[width=\columnwidth]{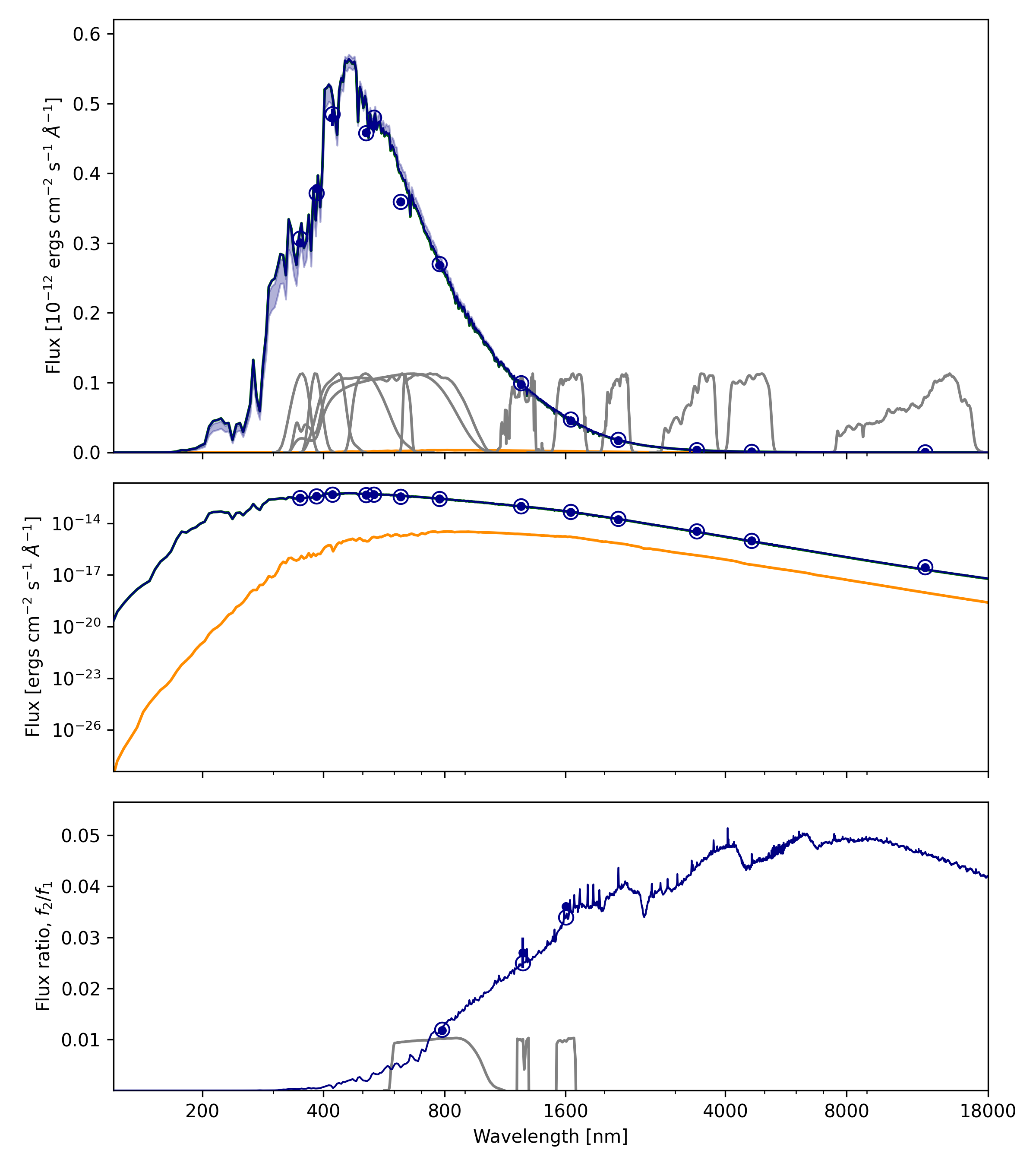}
    \caption{Upper panel: The SED of CD-27 2812. The best-fit SED is plotted as a line and
the mean SED $\pm 1-\sigma$ is plotted as a filled region. The observed fluxes
are plotted as points with error bars and predicted fluxes for the best-fit
SED integrated over the response functions shown are plotted with open
circles.  The SEDs of the two stars are also plotted. Middle panel: Same as
the upper panel but with fluxes plotted on a logarithmic scale. Lower panel:
Flux ratio as a function of wavelength for the best-fit SEDs. The observed
flux ratios are plotted as points with error bars and the predicted flux
ratios integrated over the filter  profiles shown are plotted as open circles.}
    \label{fig:sed}
\end{figure}

The posterior probability distribution for the model parameters from our analysis to measure the stellar effective temperatures is summarised in Table~\ref{tab:teb} and the spectral energy distribution is plotted in Fig.~\ref{fig:sed}.
The random errors quoted in Table~\ref{tab:teb} do not allow for the systematic error due to the uncertainty in the absolute calibration of the CALSPEC flux scale \citep{2014PASP..126..711B}. This additional systematic error is 9\,K for the F9 V primary star and 5\,K for the M-dwarf companion. 

We extracted 10\,000 values of T$_{\rm eff,1}$ and T$_{\rm eff,2}$ sampled from the PPD and paired these with the random sample of $R_1$ and $R_2$ values computed in Section~\ref{sec:mr} to measure the luminosity of the two stars with their standard errors given in Table~\ref{tab:mr}.
The two stars compared to other stars in eclipsing binary systems with accurate mass and radius measurements are shown in the Hertzsprung-Russell diagrams in Fig.~\ref{fig:hrd}.

\begin{table}
\caption{Results from our analysis to measure the effective temperatures for both stars in
CD-27 2812. The output parameter values are calculated using the mean and standard error of the
posterior probability distribution sampled using {\tt emcee}. Note that the results for star 2
are very dependent on the flux ratio priors calculated from our assumed colour\,--\,T$_{\rm
eff}$ relations, and so may be subject to additional systematic error.
$\alpha_{\rm m}$ and $\alpha_{\rm r}$ are the widths of the exponential priors $\sigma_{\rm m} $ and $\sigma_{\rm r} $, respectively. }
\label{tab:teb}
\centering
\begin{tabular}{@{}lr} 
\hline
Parameter & \multicolumn{1}{l}{Value} \\ 
\hline
\noalign{\smallskip}
\multicolumn{2}{@{}l}{Priors} \\
$\theta_1$ [mas] & $0.08375 \pm 0.00069$ \\
$\theta_2$ [mas] & $0.02524 \pm 0.00048$ \\
E(B$-$V) & $0.006 \pm 0.014 $ \\
$\alpha_{\rm m}$& 0.05 \\
$\alpha_{\rm r}$& 1e-05 \\
\noalign{\smallskip}
\multicolumn{2}{@{}l}{Model parameters} \\
$T_{\rm eff,1}$ [K] & $6197 \pm 46$ (rnd) $\pm 9$ (sys) \\
$T_{\rm eff,2}$ [K] & $3770 \pm 23$ (rnd) $\pm 5$ (sys) \\
$\theta_1$ [mas] & $0.08274 \pm 0.00076 $ \\
$\theta_2$ [mas] & $0.02426 \pm 0.00036 $ \\
E(B$-$V) &$ 0.010 \pm 0.010 $ \\
$\sigma_{\rm m} $&$ 0.012 \pm 0.010 $ \\
$\sigma_{\rm r} $&$ 0.000 \pm 0.000 $ \\
$c_{1,1} $&$ -0.173 \pm 0.184 $ \\
$c_{2,1} $&$ -0.064 \pm 0.171 $ \\
$c_{1,2} $&$  0.380 \pm 0.329 $ \\
$c_{2,2} $&$ -0.057 \pm 0.162 $ \\
$c_{1,3} $&$ -0.083 \pm 0.134 $ \\
$c_{2,3} $&$  0.120 \pm 0.155 $ \\
$c_{1,4} $&$  0.160 \pm 0.193 $ \\
$c_{2,4} $&$  0.052 \pm 0.115 $ \\
\noalign{\smallskip}
\multicolumn{2}{@{}l}{Derived parameters} \\
${\mathcal F}_{\oplus,1}$[$10^{-8}$ erg\,cm$^{-2}$\,s$^{-1}$] & $0.3364 \pm 0.0097$ (rnd) $\pm 0.0021$ (sys)  \\
${\mathcal F}_{\oplus,2}$[$10^{-10}$ erg\,cm$^{-2}$\,s$^{-1}$] & $0.3963 \pm 0.0083$ (rnd) $\pm 0.0023$ (sys)  \\
$\log(L_1/L_{\odot})$  & $0.597 \pm 0.013$ (rnd) $\pm 0.003$ (sys)  \\
$\log(L_2/L_{\odot})$  & $-1.3322 \pm 0.0097$ (rnd) $\pm 0.0025$ (sys)  \\
\noalign{\smallskip}
\hline
\end{tabular}
\end{table}

\begin{table}
\caption[]{Fundamental parameters of the stars in CD$-$27 2812.}

\label{tab:mr}
\begin{center}
  \begin{tabular}{lrrr}
\hline
\noalign{\smallskip}
 \multicolumn{1}{l}{Parameter} &
 \multicolumn{1}{l}{Value} &
 \multicolumn{1}{l}{Error} &
 \multicolumn{1}{r}{} \\
\noalign{\smallskip}
\hline
\noalign{\smallskip}
$M_1/{\mathcal M^{\rm N}_{\odot}}$&1.3597  & $\pm$ 0.0024 & [0.18 \%] \\
\noalign{\smallskip}
$M_2/{\mathcal M^{\rm N}_{\odot}}$&0.5624 & $\pm$ 0.0006 &[0.11 \%] \\
\noalign{\smallskip}
$M_2/M_1$ & $ 0.4136 $ &$ \pm 0.0003$ &  [0.07 \%] \\
\noalign{\smallskip}
$R_1/{\mathcal R^{\rm N}_{\odot}}$&1.7207 & $\pm$ 0.0041 &[0.24 \%] \\
\noalign{\smallskip}
$R_2/{\mathcal R^{\rm N}_{\odot}}$&0.5307 & $\pm$ 0.0015 &[0.28 \%] \\
\noalign{\smallskip}
$\log(T_{\rm eff,1}/{\rm K}) $  & $ 3.792 $ & $\pm$ 0.003  & [0.73 \%] \\
\noalign{\smallskip}
$\log(T_{\rm eff,2}/{\rm K)}$  & $ 3.576$ & $\pm$ 0.003  & [0.60\%] \\
\noalign{\smallskip}
$\rho_1/{\rho^{\rm N}_{\odot}}$  &0.2669 & $\pm$ 0.0019 &[0.69 \%] \\
\noalign{\smallskip}
$\rho_2/{\rho^{\rm N}_{\odot}}$  & 3.7609 & $\pm$ 0.0314 &[0.84 \%] \\
\noalign{\smallskip}
$\log g_1$ [cgs] & 4.100 & $\pm$ 0.002 & [0.47 \%]  \\
\noalign{\smallskip}
$\log g_2$ [cgs] & 4.7381 & $\pm$ 0.0024 & [0.56 \%] \\
\noalign{\smallskip}
$\log L_1/{\mathcal L^{\rm N}_{\odot}}$   & 0.595 & $\pm$ 0.013 & [2.98 \%]  \\
\noalign{\smallskip}
$\log L_2/{\mathcal L^{\rm N}_{\odot}}$   & $ -1.290 $& $\pm$ 0.011 & [2.45 \%]  \\
\noalign{\smallskip}
[Fe/H] & $0.15$ & $\pm$ 0.15 & \\ 
\noalign{\smallskip}
\hline
\end{tabular}
\end{center}
\end{table}

\section{Discussion}
\label{sec:discuss}

We compared  the parameters of the primary star, CD$-$27~2812\,A to a grid of stellar models computed with the {\sc garstec} stellar evolution code \citep{2013MNRAS.429.3645S,2008Ap&SS.316...99W} using the software package \software{bagemass} version 1.3 \citep{2015A&A...575A..36M} as described in \citep{2025MNRAS.544.4611M}.

We find a good fit to the observed properties of CD$-$27~2812\,A for models with an age of $2.71 \pm 0.06$\,Gyr assuming a mixing length parameter $\alpha_{\rm MLT}=1.5$  and solar helium abundance.
The quality of the fit is very similar for models with solar mixing length  ($\alpha_{\rm MLT}=1.78$) at an age of $3.07 \pm 0.09$\,Gyr. 
The quality of the fit for an enhanced helium abundance ($\Delta Y = 0.02$) is significantly worse.

\begin{figure}
    \centering
    \includegraphics[width=0.98\linewidth]{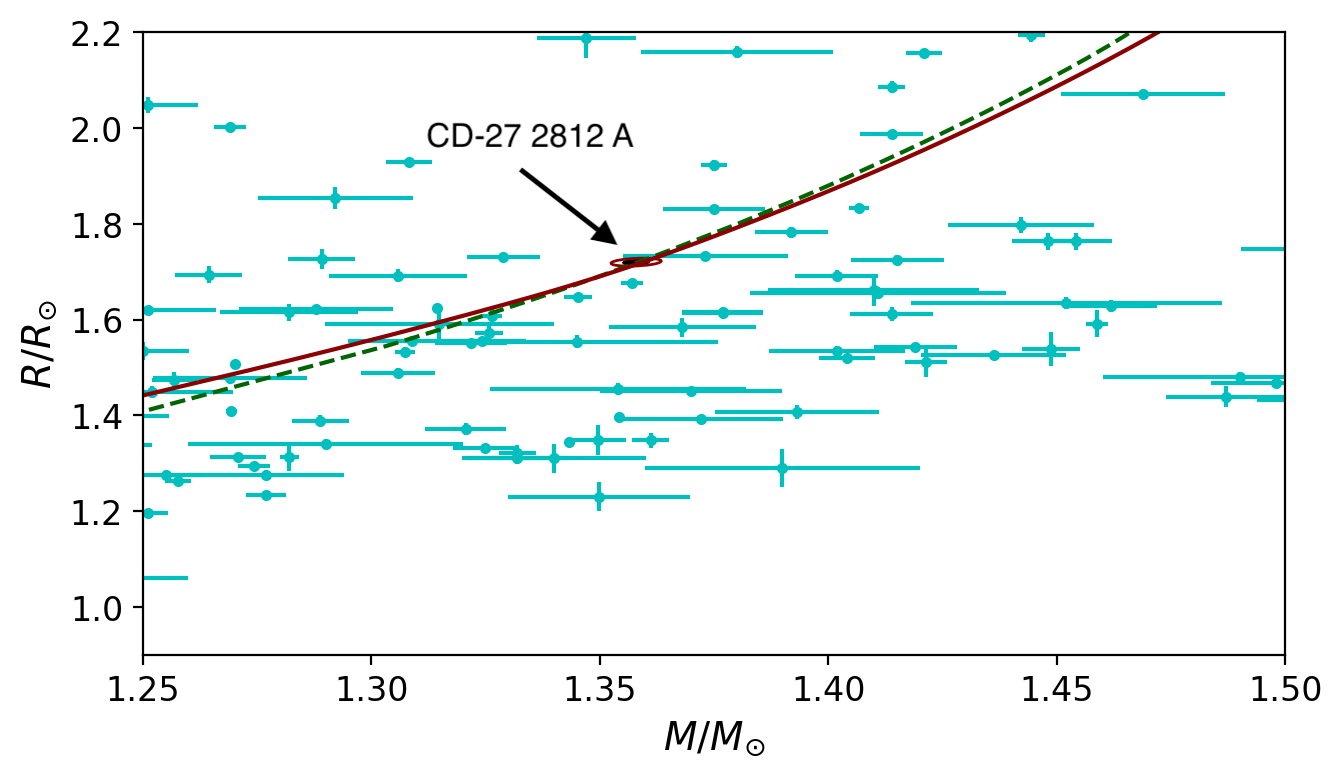}
    \includegraphics[width=0.98\linewidth]{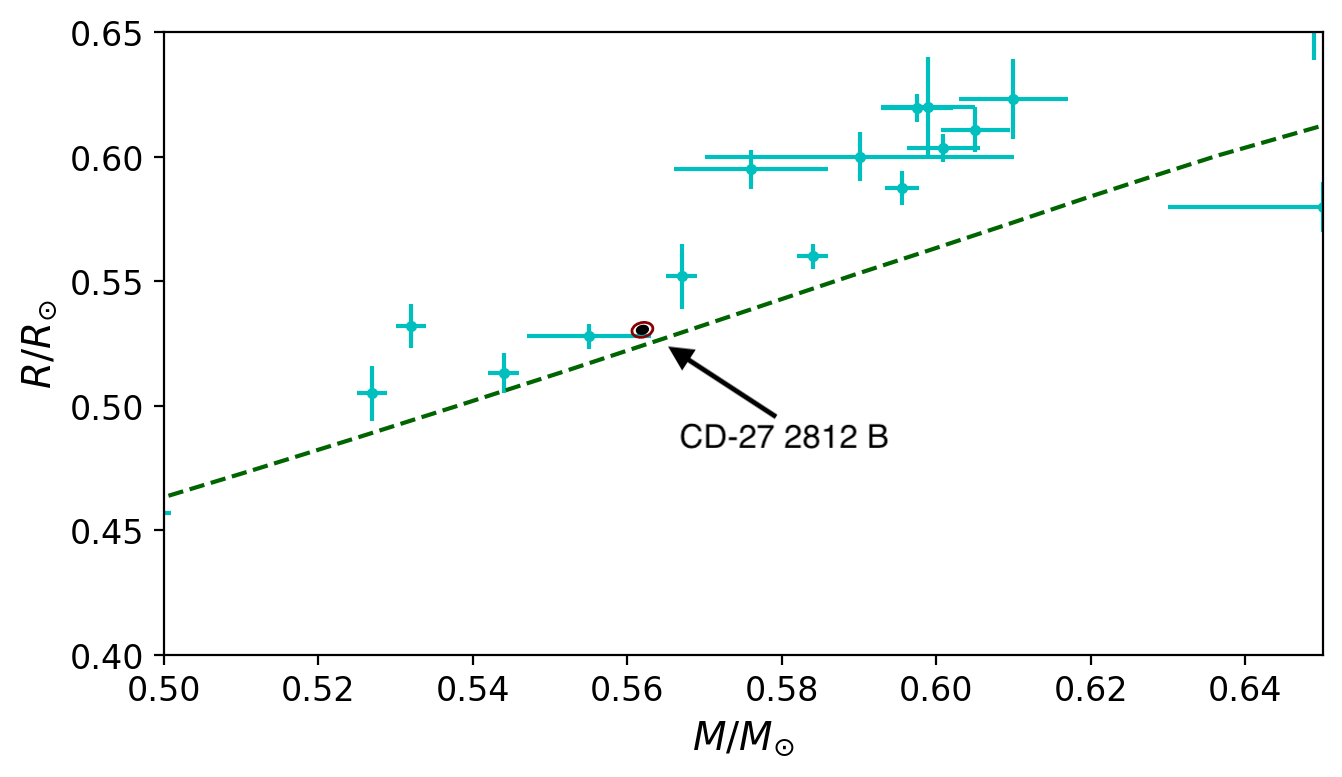}
    \includegraphics[width=0.98\linewidth]{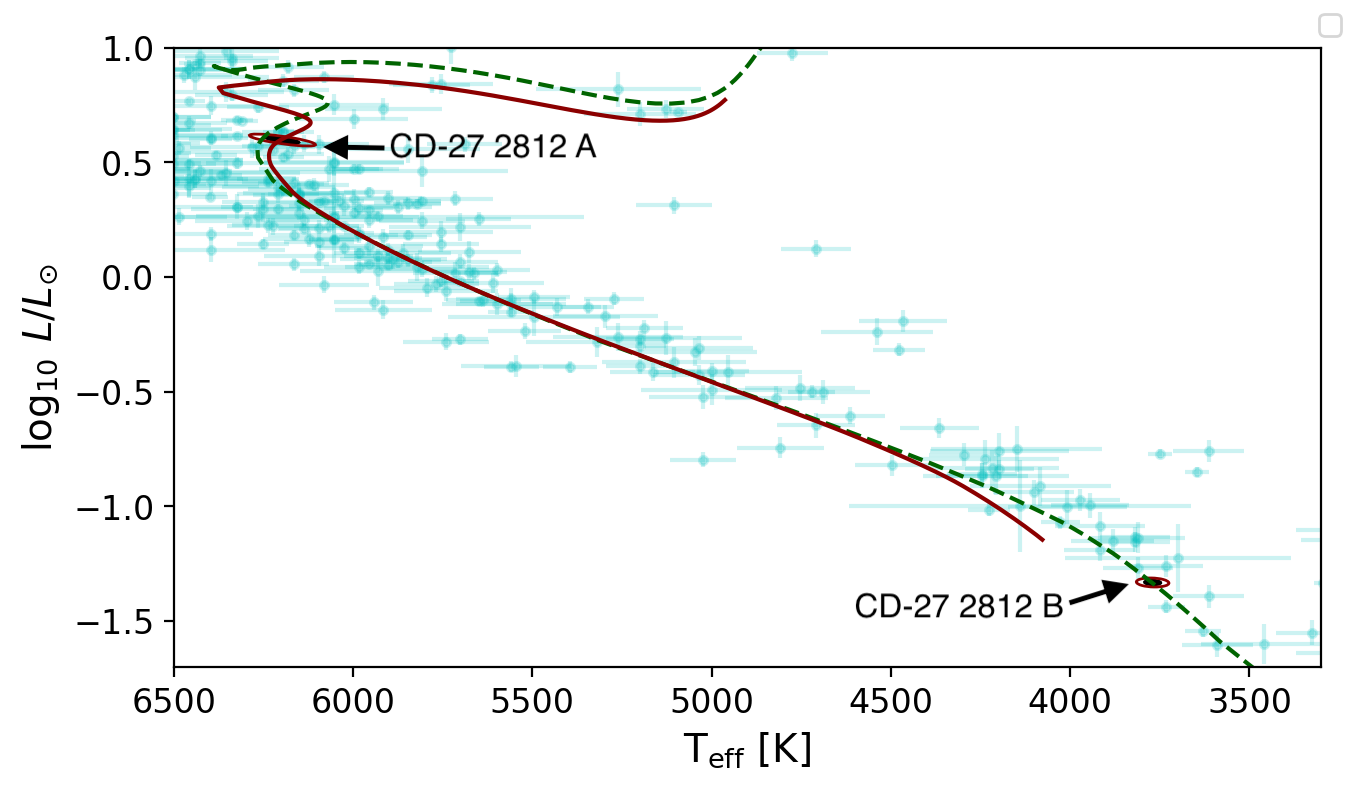}
    \includegraphics[width=0.98\linewidth]{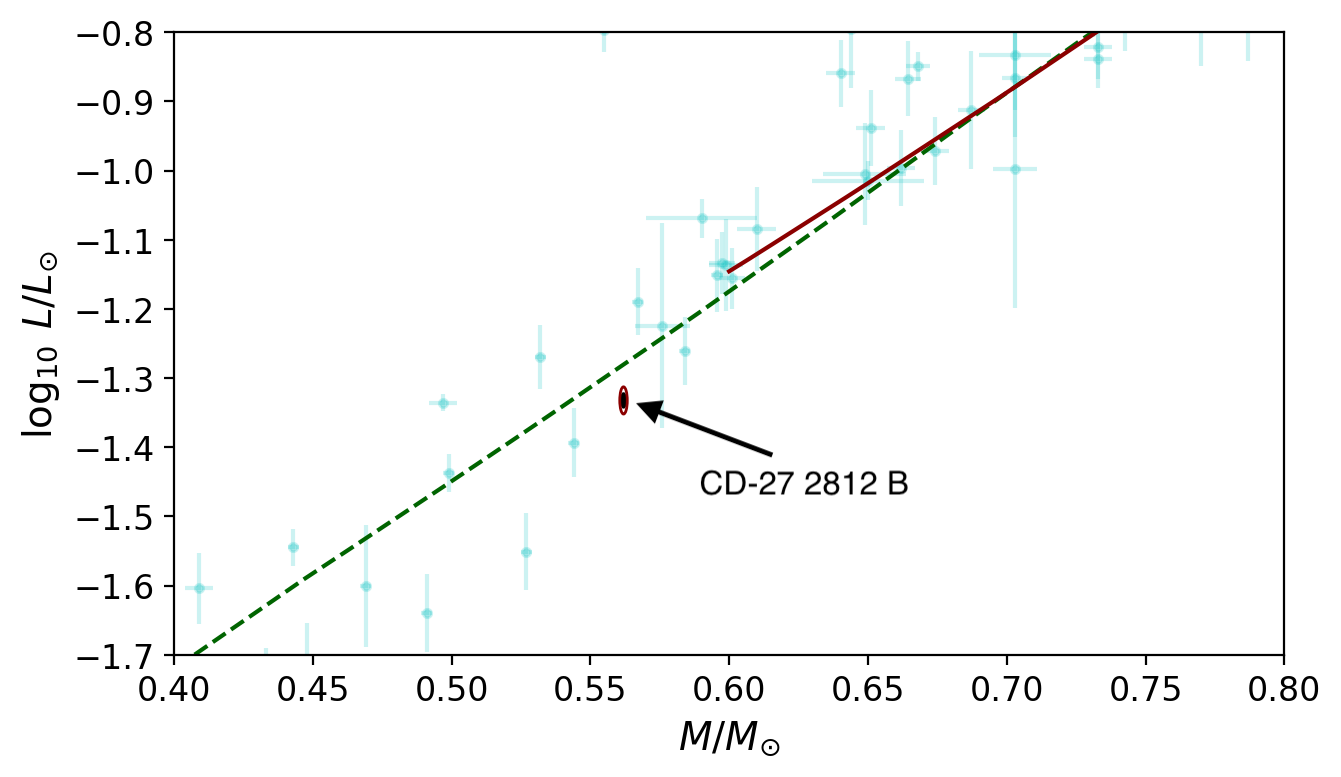}
    \caption{ Upper panel: primary component of CD$-$27~2812 in the mass\,--\,radius plane. 
    Upper Middle panel: secondary component of CD$-$27~2812 in the mass\,--\,radius plane. 
    Lower Middle panel: both components of CD$-$27~2812 in the  Hertzsprung-Russell diagram.
    Lower panel: secondary components of CD$-$27~2812 in the luminosity\,--\,mass plane.
    The ellipses show 1-$\sigma$ and  2-$\sigma$ confidence regions on the parameters of CD$-$27~2812. 
    All panels show isochrones for an age of  2.71\,Gyr assuming an initial metallicity [Fe/H]=0.164 and $\alpha_{\rm MLT}=1.5$ from \software{bagemass} (red solid line) and MIST at 2.82\,Gyr (green dashed line) for the same initial metallicity. 
    Cyan error bars show stars in eclipsing binary systems taken from DEBCat \citep{2015ASPC..496..164S}.}
    \label{fig:hrd}
\end{figure}

The best-fit isochrone with an age of 2.71\,Gyr assuming $\alpha_{\rm MLT}=1.5$ is shown in Fig.~\ref{fig:hrd}.
Isochrones for the same age and initial metal abundance from the MESA Isochrones \& Stellar Tracks \citep[MIST, ][]{2016ApJ...823..102C} are also shown in Fig.~\ref{fig:hrd}.

CD$-$27~2812\,B matches very well the mass-radius relation for low-mass stars from the MIST stellar model grids interpolated to the age and and metallicity of the binary systems estimates from the properties of the primary component (Fig. ~\ref{fig:hrd}). This is unusual for stars of this mass range, which tend to be larger than predicted by stellar models. These stars also tend to be cooler than predicted \citep{2013ApJ...776...87S}.   

\cite{2023MNRAS.522.2683M} derived the following masses and radii for the stars in CD$-$27~2812: $M_1 = 1.19 \pm 0.08 \,M_{\odot}$,  $R_1 = 1.66 \pm 0,04\,R_{\odot}$, $M_2 = 0.51 \pm 0.02\,M_{\odot}$, $R_2 = 0.51 \pm 0.01 \,R_{\odot}$. 
The value of K$_2$ that we have used was not available at the time of that study so constraints on the primary star's radius and mass derived from the analysis of its spectral energy distribution and spectrum were required to obtain these mass and radius estimates.
If we use the values of K$_1$, T$_{\rm eff,1}$ and [Fe/H] from this study and re-calculate the primary star mass using the same method as  \cite{2023MNRAS.522.2683M} we obtain the value $M_1 = 1.24 \pm 0.09\,M_{\odot}$. 
This improved agreement is mostly due to the change in the estimate of [Fe/H].
The better precision in the values we have derived is mostly due to our access to observations from HARPS and NIRPS.
This has enabled us to directly and precisely measure the masses for both stars.

The observational techniques we have developed here can be applied to other eclipsing M-dwarf systems to provide valuable information for testing stellar atmosphere models of M-dwarfs.  
The need for these observational tests can be seen in Figure~1 of \cite{Iyer_2023} where variations $\approx$10\,per~cent can be seen in the spectral shape of M-dwarfs in the near infrared for different models.
We saw the impact of these differences between models during our analysis of CD$-$27~2812. 
We found during the subtraction of the synthetic M-dwarf spectra that the flux ratios calculated for the near infrared bands were notably different depending on which stellar model was used to create the template. 
We found that the Bt-Settl and Bt-NEXTGEN models produced flux ratios with values that we expected, whereas the Phoenix models produced ratios that were up to 0.01 too low compared to the expected values based on the T$_{\rm eff}$-colour relations for M-dwarf. When the Phoenix model fluxes were used in {\sc teb} we found that the produced chi-squared values were, at minimum, double than what was produced for this paper, additionally the Phoenix fluxes would also produce a $T_{\rm eff,2}$ $\sim$ 100K lower than expected.
The  interpretation of these results is not straightforward because the amplitude of the peak due to the M-dwarf in stacked CCF is determined by both the total flux in the band from the M-dwarf and the typical line strength in the model spectra.
Direct measurements of the flux ratio from infrared light curves covering the eclipse of the M-dwarf would be very helpful to resolve this ambiguity.

\section{Conclusion}
\label{sec:conclusions}
CD$-$27~2812 adds to a small but growing sample of eclipsing binaries for which we have directly measured the effective temperature of a solar-type star with a much fainter M-dwarf companion \citep{2022MNRAS.513.6042M, 2023MNRAS.522.2683M,2024MNRAS.531.4577M,2025MNRAS.544.4611M}. 
These stars are complementary to benchmark stars for which the effective temperature estimate is based on the angular diameters ($\theta$) measured using interferometry. 
The effective temperature measurements we have made from eclipsing binary stars have the advantage that the sources of systematic error are small and well-understood, and the surface gravity of the stars is known to very high accuracy. 
In addition, these stars are within the magnitude range (V$\sim 10$) that can be observed directly by instruments used for large-scale spectroscopic surveys in their standard observing modes. 
This makes it feasible to conduct ``end-to-end'' tests of the accuracy of stellar parameters derived by these surveys.
This is often not the case for the very bright stars that are currently accessible to interferometric measurements. 

We have been able to directly measure the fundamental properties of the M-dwarf star in this eclipsing binary. 
In doing so we have found variations in stellar atmospheric models currently available. 
Additionally, having an M-dwarf with its parameters known to an accurate degree can contribute towards the improvement of stellar evolutionary tracks for low mass stars and further our understanding of the radius inflation problem. 

\section*{Acknowledgements}
We thank the anonymous referee for their careful reading of the manuscript and comment that have helped to improve the paper.

PM acknowledges support from STFC research grants ST/Y002563/1 and UKRI1193 and UK Space agency grant UKRI966.

Based on observations collected at the European Organisation for Astronomical Research in the Southern Hemisphere under ESO programme 112.25UD.

We thank Valentin Ivanov for processing the NIRPS data used in this research. 

 This work has made use of data from the European Space Agency (ESA) mission
{\it Gaia} (\url{https://www.cosmos.esa.int/gaia}), processed by the {\it Gaia}
Data Processing and Analysis Consortium (DPAC,
\url{https://www.cosmos.esa.int/web/gaia/dpac/consortium}).
Funding for the DPAC has been provided by national institutions, in particular the institutions participating in the {\it Gaia} Multilateral Agreement.

This paper includes data obtained through the TESS Guest investigator programs  G06022 (PI Martin), G05024 (PI Martin), G04157 (PI Martin), G03216 (PI Martin) and G022253 (PI Martin).

This paper includes data collected by the TESS mission, which is publicly available from the Mikulski Archive for Space Telescopes (MAST) at the Space Telescope Science Institure (STScI). Funding for the TESS mission is provided by the NASA Explorer Program directorate. STScI is operated by the Association of Universities for Research in Astronomy, Inc., under NASA contract NAS 5–26555. We acknowledge the use of public TESS Alert data from pipelines at the TESS Science Office and at the TESS Science Processing Operations Center.

This research made use of Lightkurve, a Python package for Kepler and TESS data analysis \citep{2018ascl.soft12013L}.

\section*{Data Availability}
The data underlying this article are available in the following repositories:  
Mikulski Archive for Space Telescopes -- \url{https://archive.stsci.edu/}  (TESS); ESO Science Archive Facility -- \url{https://archive.eso.org/}.



\bibliographystyle{mnras}

\bibliography{allbib} 


\appendix
\section{Simultaneous NIRPS+HARPS RV fit with jktebop}
\label{sec:jktebop}

We wanted to account for the offset between the radial velocity measurements obtained with the HARPS and NIRPS spectrographs. 
There is no straightforward way to do this in {\sc  jktebop} for different RV data sets that overlap in time. 
We were able to work around this by shifting the time stamps on the NIRPS data by an integer number of orbital cycles, $n_P$. 
By using $n_P=23$, the first shifted time stamp for the NIRPS data is greater than the last time stamp for the HARPS data.
We can then use the {\sc poly} function in the {\sc jktebop} input file to specify a new free parameter, $V_{0,{\rm NIRPS}} - V_{0,{\rm HARPS}} $, that is added to the systematic velocity of the primary star orbit for RV data with time stamps corresponding to NIRPS observations.

\bsp	
\label{lastpage}
\end{document}